\documentclass{ICEAA-IEEE_APWC}
\pdfoutput=1
\usepackage{color}
\usepackage{amsmath}
\usepackage{amsfonts}
\usepackage{graphicx}
\usepackage{caption}
\usepackage{subcaption}
\usepackage{booktabs}
\usepackage[numbers,sort]{natbib}
\usepackage{url}
\usepackage{aas_macros}
\usepackage[binary-units=true]{siunitx}

\usepackage{needspace}

\title{Interference Mitigation with a Modified ASKAP Phased Array Feed on the \SI[detect-weight]{64}{\metre} Parkes \\Radio Telescope}

\author{%
A.P. Chippendale\thanks{CSIRO Astronomy and Space Science, PO Box 76, Epping, 1710, Australia, e-mail: Aaron.Chippendale@csiro.au} \and G. Hellbourg\thanks{Department of Astronomy, University of California, 339 Campbell Hall, Berkeley, CA 94710, USA}}

\begin{document}
\maketitle

\begin{abstract}
We present results from a first attempt to mitigate radio frequency interference in real-time during astronomical measurements with a phased array feed on the \SI[detect-weight]{64}{\metre} Parkes radio telescope. Suppression of up to \SI[detect-weight]{20}{\decibel} was achieved despite errors in estimating the interference spatial signature.  Best results were achieved in the clean excision of a narrowband and stationary clock signal that originates from the receiver's digital back-end system.   We also contribute a method to interpolate valid beamformer weights at interference-affected channels.  Correct initial beam weights are required to avoid suppressing the desired signal.
\end{abstract}

\section{AN ASKAP PAF AT PARKES}
A phased array feed (PAF) is a dense array of antenna elements at the focus of a concentrator that, with digital beamforming, can produce multiple simultaneous antenna beams of high sensitivity throughout a wide field of view.
%PAF offer faster sky survey speed.
%PAFs boost the speed with which radio telescopes can survey large areas of the radio sky. % TODO: global PAF citations
We trialled interference mitigation with a Mk. II PAF system \cite{Hampson2012advancing} built by CSIRO for the Australian Square Kilometre Array Pathfinder (ASKAP) telescope 
%\cite{DeBoer2009Australian} 
and modified
%, via a collaboration between CSIRO and the Max Planck Institute for Radio Astronomy (MPIfR), for deployment on the Effelsberg 100 m telescope.
by CSIRO and the Max Planck Institute for Radio Astronomy (MPIfR) for deployment on the Effelsberg \SI{100}{\metre} telescope.
%We took data when the PAF was installed on the 64 m Parkes radio telescope for scientific commissioning.
Our experiments took place during a scientific commissioning phase on the \SI{64}{\metre} Parkes telescope \cite{Chippendale2016testing}.
%,Deng2017observing,Reynolds2017Spectral}.

%\subsection{Frontend}
The ASKAP PAF \cite{Hampson2012advancing} is a connected-element ``chequerboard'' array \cite{Hay2008analysis} operating over \SIrange[range-phrase = --,range-units = single]{0.7}{1.8}{\GHz}.  Modifications for MPIfR involved
%adding more selective filters in the PAF to reject broadband telecom in 694-820 MHz, 850-960 MHz and 1805-1880 MHz, and aeronautical ADS-B/DME at 960-1164 MHz \cite{Chippendale2016testing}.
the addition of selective filters to reject broad-band radio frequency interference (RFI) \cite{Chippendale2016testing}.
%However, in this work we use an unmodified ASKAP filter covering 700-1200 MHz that admits many of these strong interfering signals. 
The experiment described here has however used an unmodified ASKAP filter covering \SIrange[range-phrase = --,range-units = single]{0.7}{1.2}{\GHz}.

%\subsection{Digital backend}
The digital beamformer can simultaneously form 36 dual-polarisation beams, each of \SI{384}{\MHz} bandwidth, from the PAF's 188 antenna elements. The beamformer also calculates array covariance matrices (ACMs) with \SI{1}{\MHz} resolution and integrated fine filter bank (FFB) spectra of the beamformer outputs with \SI{18.5}{kHz} resolution.  

\begin{figure}[t]
% trim left lower right upper
\includegraphics[width=\columnwidth, trim=15mm 2mm 26mm 23mm, clip]{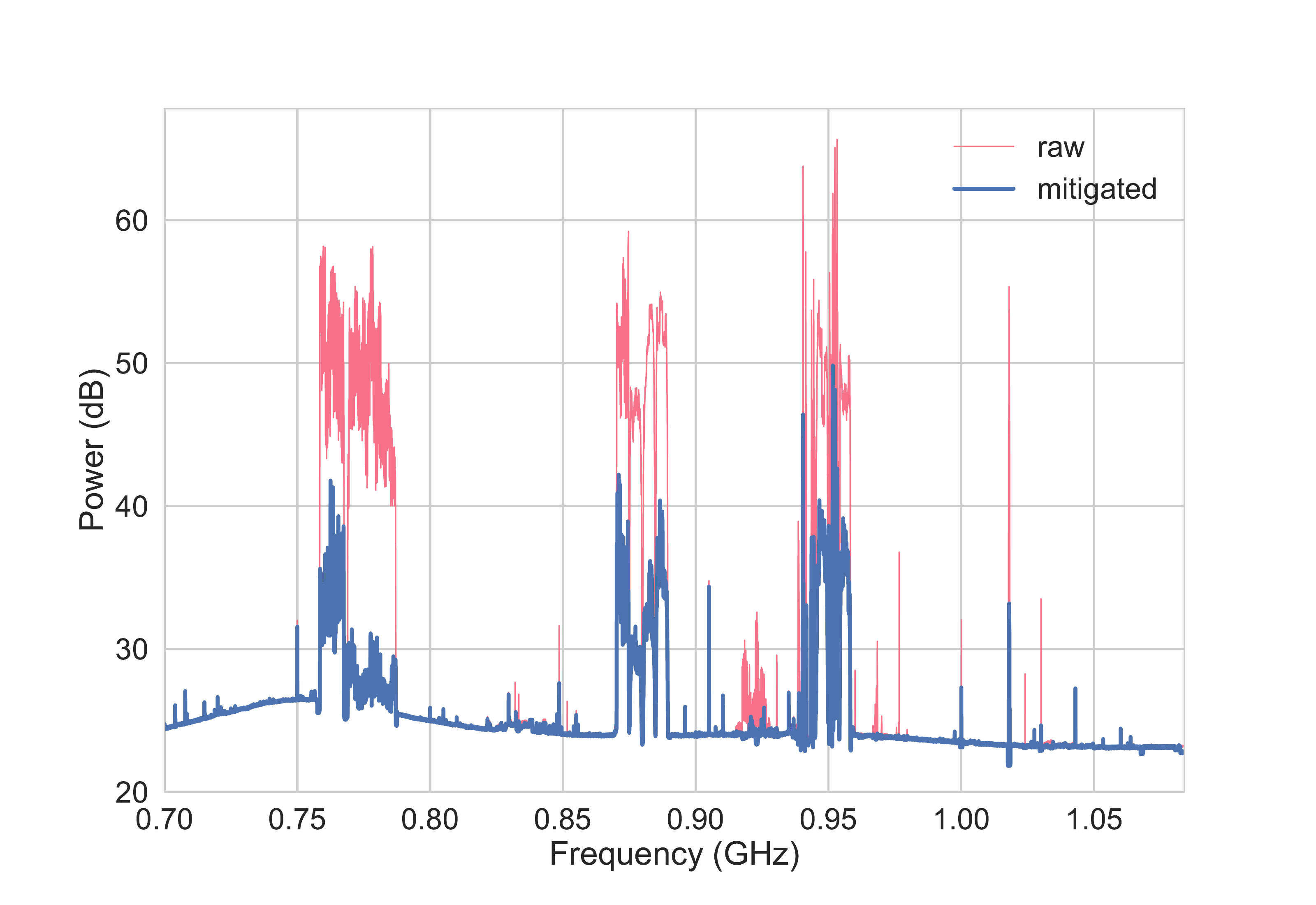} 
\caption{\small Median spectrum at Parkes with and without mitigation applied to a boresight maxSNR beam.\label{fig:spectrum} }
\end{figure}

Parkes has more challenging RFI than the radio-quiet site that the ASKAP PAF was designed for \cite{Indermuehle2016askap}. Figure \ref{fig:spectrum} shows median beamformed spectra, with and without mitigation, for maximum signal-to-noise ratio (maxSNR \cite{van1988beamforming}) weights over a \SI{13}{\minute} observation with \SI{18.5}{\kHz} resolution.
%over 700-1084 MHz.   
Table \ref{tab:rfi}
%provides identification of
lists dominant signals in this data.  RFI scans over the \SIrange[range-phrase = --,range-units = single]{0.7}{1.8}{\GHz} range of the PAF, before beamforming, are given in \cite{Chippendale2016testing}.  The broader RFI status at Parkes is summarised on the observatory's website\footnote{\url{http://www.parkes.atnf.csiro.au/observing/NEW/RFIStatus_June2016.pdf}}.

\begin{table*}
\begin{center}
\caption{Dominant RFI Identification}\label{tab:rfi}
\begin{footnotesize}
\begin{tabular}{crlcp{4.5cm}l}
\toprule
Frequency     & Bandwidth  & Service         & $n_\text{opt}$ &Comment & Label \\
%  &            &                      & & \\
(MHz)      &            &                        & & \\
\midrule
%763.000000 & 10~MHz      & 4G Mobile Telecom  & Optus &   & Tower transmitter (FDD-LTE) \\
%778.000000 & 20~MHz      & 4G Mobile Telecom & Telstra &   & Tower transmitter (FDD-LTE)  \\
773.000 & \SI{30}{\MHz}      & 4G mobile telecom   & $>10$  & Tower transmitters (FDD-LTE) & A \\ %(Optus/Telstra)
%872.500000 & 5~MHz       & 4G Mobile Telecom & Vodafone Hutchinson  &   & Tower transmitter \\
%882.500000 & 15~MHz       & 3G Mobile Telecom & Telstra  &   & Tower transmitter \\
880.000 & \SI{20}{\kHz}       & 4G mobile telecom  & $>10$  & Tower transmitters & B\\ %(Vodafone/Telstra)
%910.222222 & $<18.5$~kHz & Clock ($3^\text{rd}$ harmonic)               & CSIRO &  & seen at 910.235046387~MHz  \\
947.500 & \SI{12.5}{\MHz}     & 2/3/4G mobile telecom   & $>10$   & Tower transmitters (LTE/GSM) & C \\%, many 200 kHz GSM channels and some broader LTE \\ %(Telstra/Optus/Vodafone)
%967.600000 & 200~kHz      & L-DACS2?           & & & Digital aeronautical comms \\
%968.000000 & 200~kHz      & L-DACS2?           & & & Digital aeronautical comms\\
%968.400000 & 200~kHz      & L-DACS2?           & & & Digital aeronautical comms\\
976.593 & \SI{<18.5}{\kHz} & Clock               (CSIRO) & 1 & Seen at \SI{976.609}{\MHz} & D \\
1018.000 & \SI{500}{\kHz}    & Aeronautical radiolocation & 4 & Parkes VOR/DME transponder & E\\
%1042.962963 & $<18.5$~kHz & Clock ($5^\text{th}$ harmonic)               & CSIRO &  & seen at 1042.96398926~MHz  \\

\bottomrule
\end{tabular}
\end{footnotesize}
\end{center}
\end{table*}

\section{SPATIAL MITIGATION WITH PAFS}
Blind spatial filtering is proposed to perform RFI mitigation. The implemented approach involves (1) estimating the interfering signal subspace from the ACM, and (2) updating the beamformer weights to create spatial nulls towards the RFI. The viability of this technique has been demonstrated via simulation for full ASKAP \cite{Black2015multi} and by an experiment on ASKAP's six-antenna Boolardy Engineering Test Array (BETA) \cite{7833528}. This experiment successfully imaged a celestial source in a \SI{1}{\MHz} channel that was otherwise saturated by interference from GPS L2 over 200 times stronger than the desired signal.    

Here we trial spatial filtering across the full \SI{384}{\MHz} system bandwidth and evaluate its impact on \SI{18.5}{\kHz} resolution spectra. We explore the efficacy of the technique on a wider variety of RFI signals and measure its impact on spectral-line observations of a celestial source.   

\subsection{Experimental method at Parkes}
We used FFB spectra to assess the performance of various mitigation algorithms and their parameters. Each beam was initiated with maxSNR boresight beam weights \cite{McConnell2016Australian} with different real-time RFI mitigation algorithms and parameters. The first beam remained unprocessed for reference.  Both orthogonal and oblique projectors with various RFI subspace dimension were trialled as in \cite{7833528}.  The subspace dimension is closely related to the number of RFI sources, their bandwidths, and their relative motion to the telescope \cite{Hellbourg2015Subspace}.

ACMs were continuously downloaded to estimate the RFI subspace and compute the spatial filter, before updating beam weights. The system completes an ACM download and beam weights upload every \SI{15}{\second}.  Eight cycles of ACM downloads and weights uploads were required to apply mitigation to the full \SI{384}{\MHz} bandwidth because the ACM calculation operates on \SI{48}{\MHz} at a time. This resulted in a \SI{120}{\second} update period for any given \SI{1}{\MHz} channel. This slow rate, together with the five-fold reduction in beam feature size for the \SI{64}{\metre} Parkes antenna with respect to the \SI{12}{\metre} ASKAP antennas, affects the attenuation of RFI sources moving relatively to the telescope. The update rate could, however, be increased significantly by refining software.

The user-defined \textit{on-the-sky} ACM integration time was \SI{0.5}{\second}, resulting in a \SI{2}{\second} accumulation period as the ACM calculation in firmware uses every fourth sample.  The \SI{18.5}{\kHz} FFB spectra were integrated for \SI{2}{\second} and were downloaded every \SI{4.5}{\second}.

The experiments ran over two \SI{384}{\MHz} bands centred at \SI{891.5}{\MHz} and \SI{1340.5}{\MHz}.  At each band, we made measurements both with the antenna stationary and tracking a celestial source at the sidereal rate of \SI[per-mode=symbol]{0.25}{\degree\per\minute}.  We tracked flux calibrator PKS\;B\num[minimum-integer-digits = 4]{0407}\num{-658} \cite{Otrupcek1990Parkes} and a reference position offset by \SI[retain-explicit-plus]{+10}{\arcminute} in right ascension. Only results of the band centred at \SI{891.5}{\MHz} are presented as they contain more fixed terrestrial sources of RFI.

\section{RESULTS}
\subsection{Suppression and subspace dimension}

% trim left lower right upper
%\begin{figure}[tb]
% \centerline{\includegraphics[width=\columnwidth, trim=20mm 20mm 34mm 32mm, clip]{figures/suppression_vs_dim_exp_5}} \caption{\small Median suppression achieved by oblique projection versus RFI subspace dimension $n$ with stationary antenna.\label{fig:sup_vs_dim} }
%\end{figure}
\begin{figure}[tb]
 \centerline{\includegraphics[width=0.9\columnwidth, trim=18mm 10mm 28mm 22mm, clip]{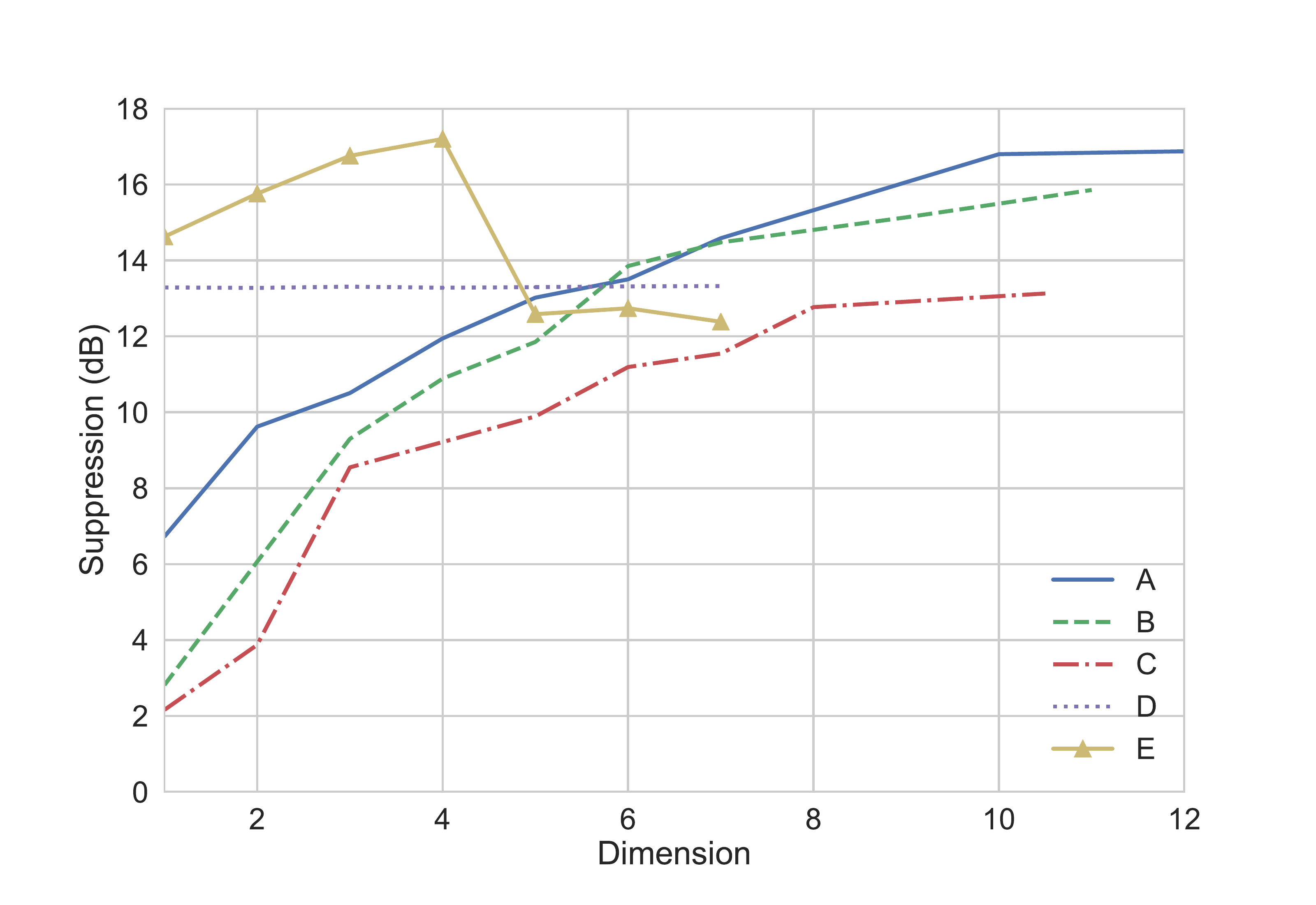}} \caption{\small Median suppression achieved over \SI{4}{\minute} by oblique projection versus RFI subspace dimension with stationary antenna.  Labels match signals to Table \ref{tab:rfi}. }\label{fig:sup_vs_dim}
\end{figure}

Figure \ref{fig:sup_vs_dim} shows RFI attenuation as a function of the dimension $n$ of the projected-out RFI subspace. This plots the ratio in dB of the total power, in each signal sub-band of Table \ref{tab:rfi}, between a beam with no mitigation and a beam with oblique projection mitigation.  Only data from when the antenna was stationary were used.  The distance measuring equipment (DME) signal (E) at \SI{1018}{\MHz} achieves peak suppression at $n=4$ and the narrowband clock signal (D) does so at $n=1$.  Suppression of the mobile telecommunication base-station signals (A-C) continues to rise at $n=10$.

% trim left lower right upper
\begin{figure}[tb]
 \centerline{\includegraphics[width=\columnwidth, trim=20mm 10mm 44mm 25mm, clip]{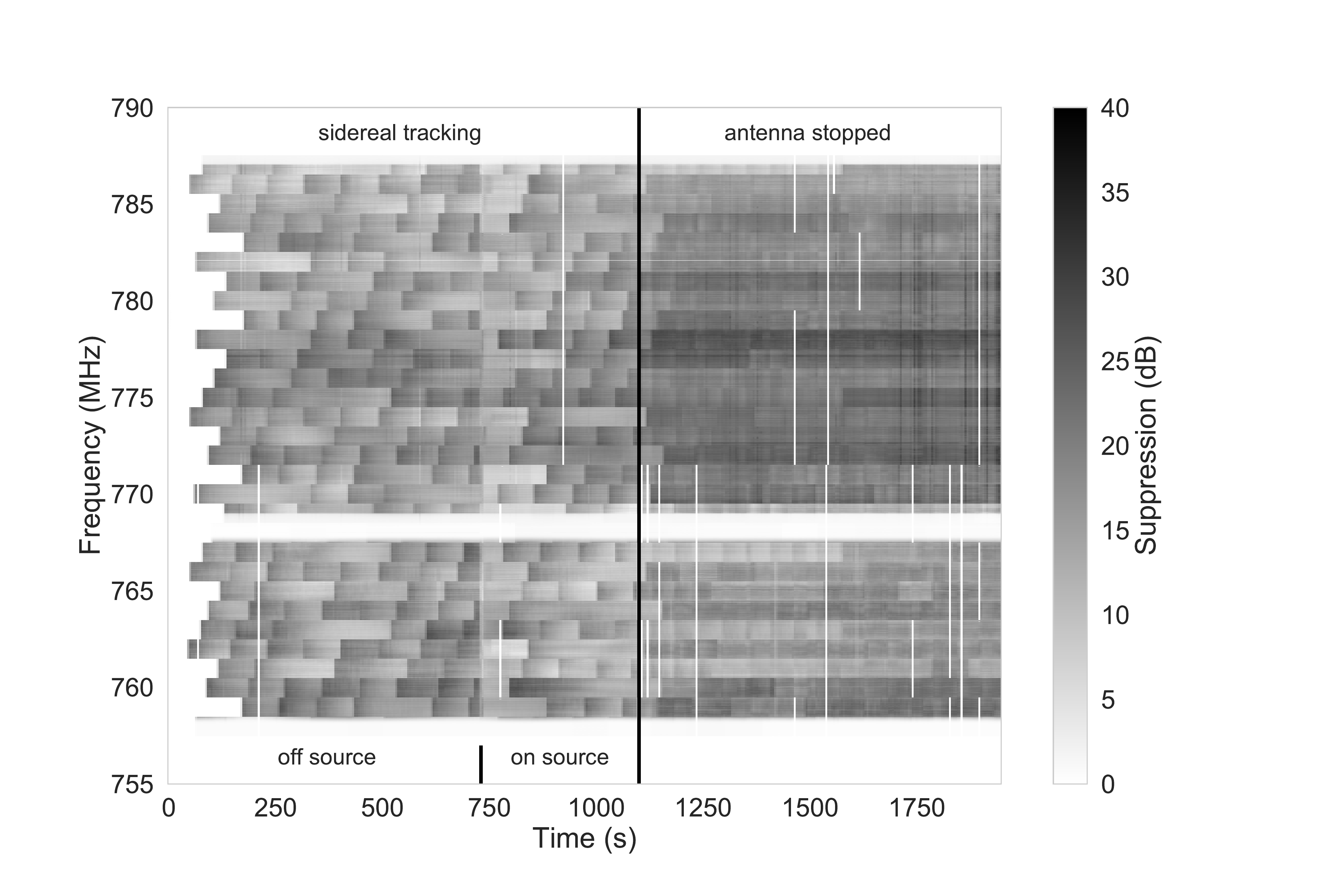}} \caption{\small Suppression  versus telescope motion for 4G mobile telecom (FDD-LTE) base-station signals.\label{fig:sup_vs_motion} }
 % (experiment E).
\end{figure}

Figure \ref{fig:sup_vs_motion} shows the change in RFI suppression between the tracking and stationary telescope. The level of suppression is more consistent when the telescope is stationary. Error in estimating the RFI subspace also limits suppression.  Performance can be improved by using reference antennas tracking RFI sources to more accurately estimate their subspace \cite{Hellbourg2014Reference}, reducing ACM integration time to reduce subspace smearing \cite{Hellbourg2015Subspace} caused by relative motion between RFI sources and the telescope, and by increasing beam weight update rates so that the mitigation is still valid when applied.

\subsection{Clean excision of a narrowband signal}
Expecting the current implementation to work best on stationary signals, we explored performance on a tone generated by the ASKAP digital receiver that is narrower than an \SI{18.5}{\kHz} FFB channel. The source is a \SI{256}{\MHz} FPGA clock that is multiplied by 32/27 to read out the digital receiver's \SI{1}{\MHz} resolution oversampled coarse filter bank \cite{7369562}.% Harmonics of this clock occur at integer multiples of $256\times32/27$~MHz and are   

% trim left lower right upper
\begin{figure}[tb]
 \centerline{\includegraphics[width=\columnwidth, trim=10mm 0mm 10mm 10mm, clip]{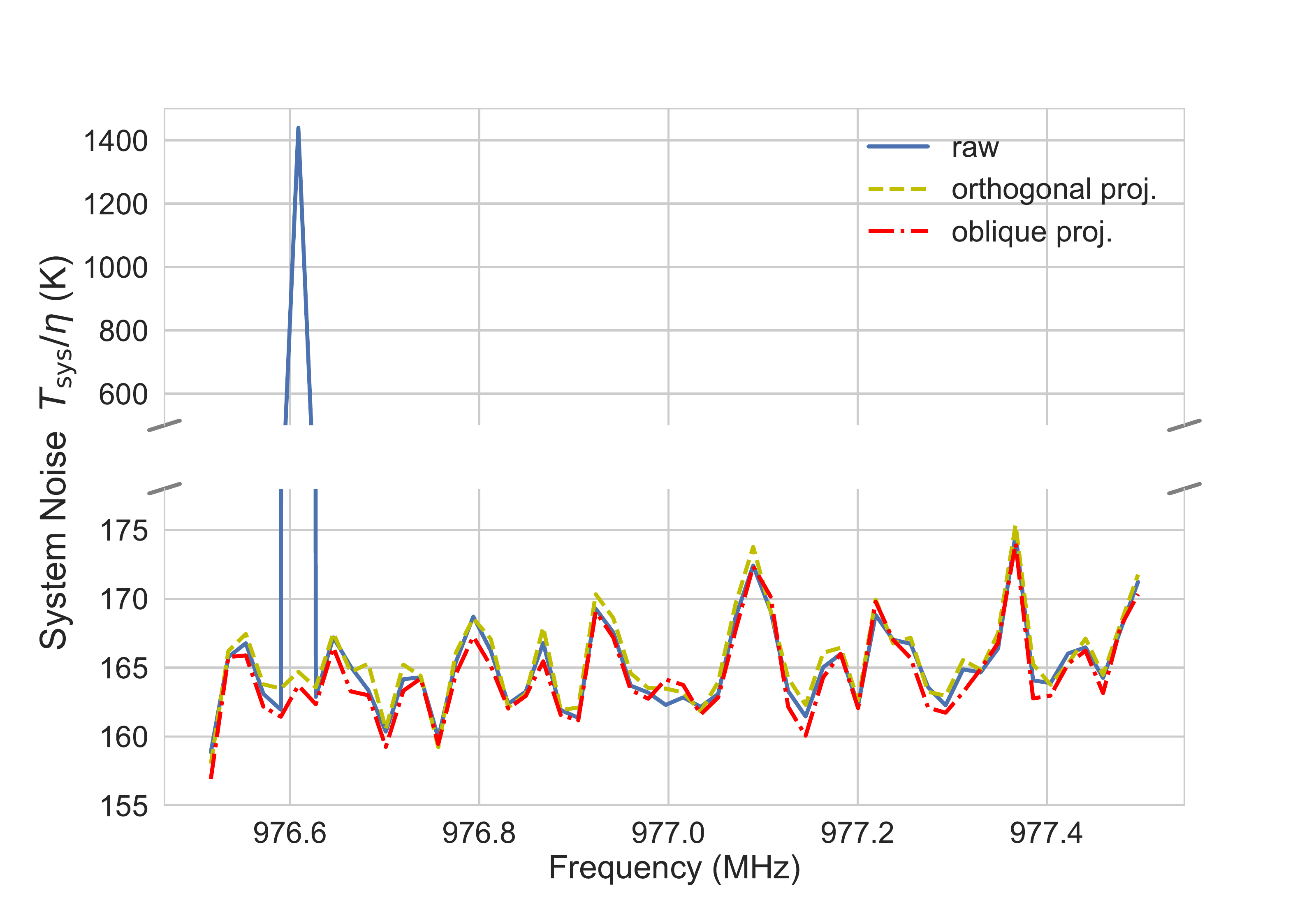}} \caption{\small Clean removal of coarse filter bank read-out clock with negligible system temperature increase.\label{fig:sensitivity} }
\end{figure}

Figure \ref{fig:sensitivity} shows that this signal, which appears at \SI{976.6}{\MHz} due to direct sampling in the second Nyquist zone, is neatly mitigated to the noise floor by both orthogonal and oblique projection with ${n=1}$.  This plot shows the beam equivalent system-temperature-on-efficiency $T_\text{sys}/\eta$ measured on PKS\;B\num[minimum-integer-digits = 4]{0407}\num{-658} assuming a flux model\footnote{\url{https://www.sao.ru/cats/cgi-bin/pkscat90.cgi}} fit to data in the Parkes catalogue \cite{Otrupcek1990Parkes}. %The lower panel zooms in to show that 
Both mitigation algorithms change system temperature by less than \SI{1}{\kelvin} with respect to the unmitigated spectrum. The typical system temperature here is higher than reported in \citep{Chippendale2016testing}, likely because we initialised the system with 35 day-old beam weights made at different azimuth and elevation and without adjustment for changes in system status since the weights were last calibrated.

\subsection{Interpolating correct initial weights}
\begin{figure}
\begin{subfigure}[tb]{\columnwidth}
\centerline{\includegraphics[width=0.93\columnwidth, trim=10mm 0mm 24mm 13mm, clip]{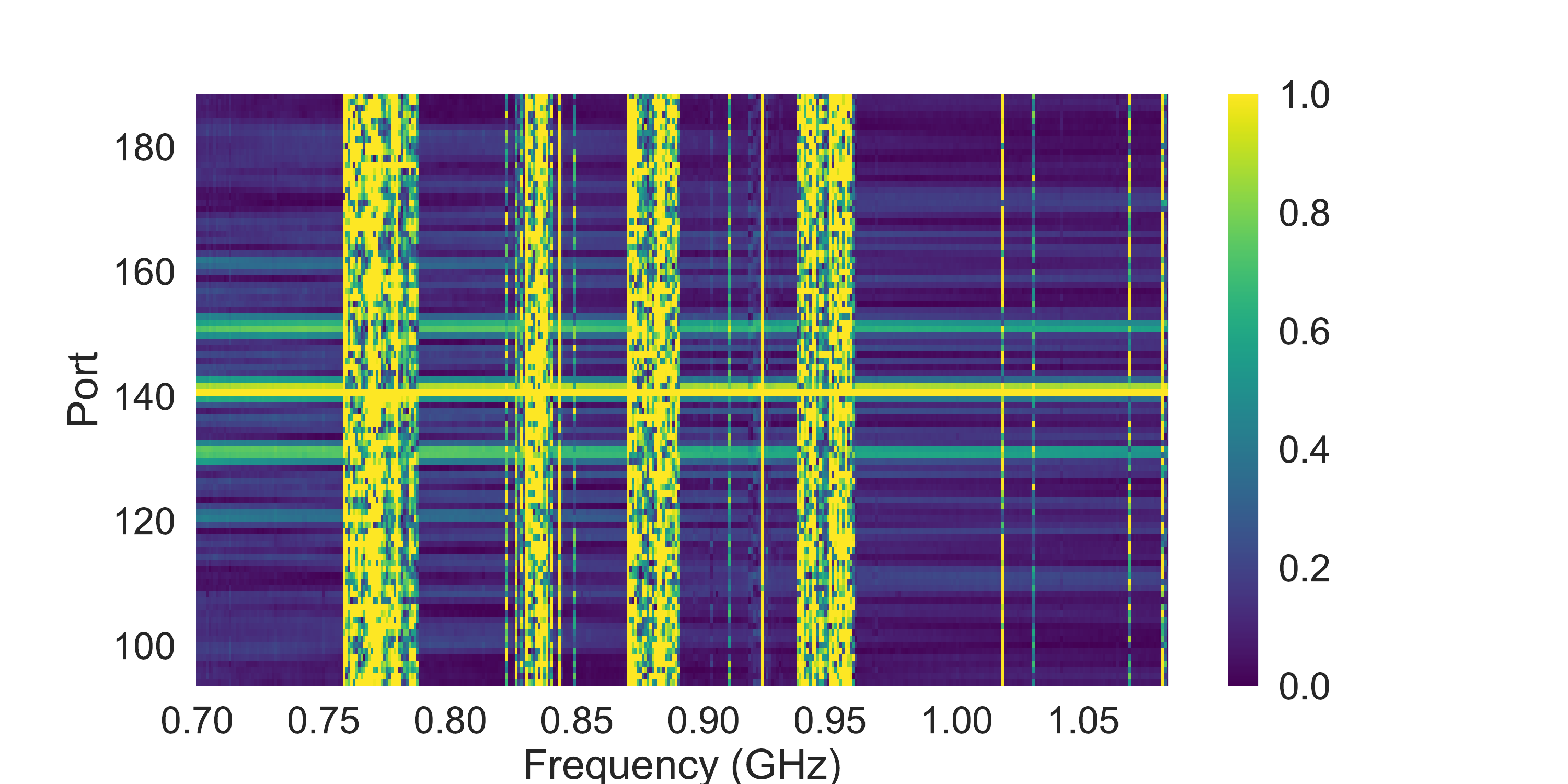}}
\caption{\small Raw.\label{fig:amp} }
\end{subfigure}

\begin{subfigure}[tb]{\columnwidth}
\centerline{\includegraphics[width=0.93\columnwidth, trim=10mm 0mm 24mm 13mm, clip]{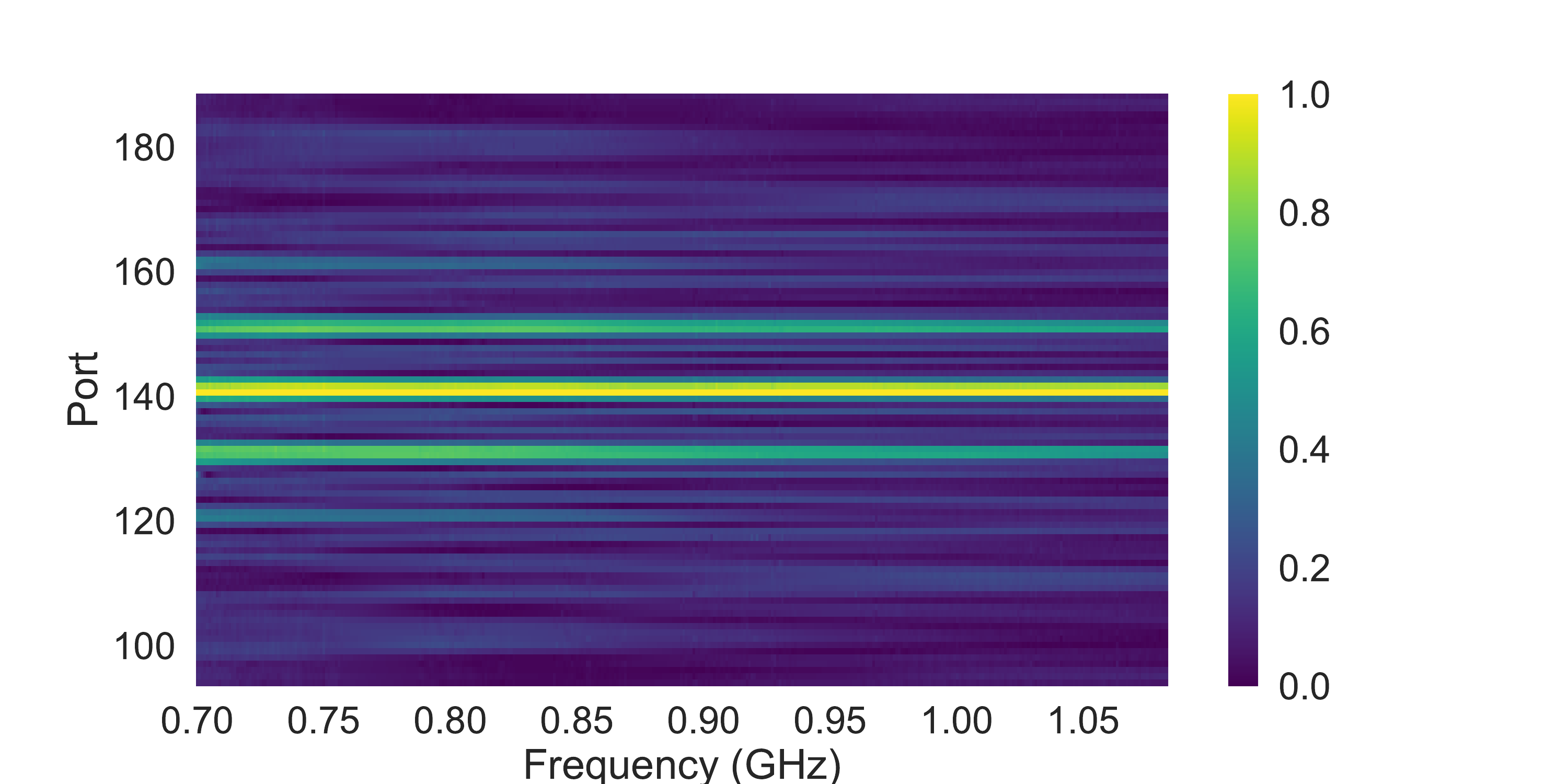}} 
\caption{\small Interpolated. \label{fig:amp-interp} }
 \end{subfigure}
 \caption{MaxSNR beam weight amplitudes. }
 \end{figure}
 
 \begin{figure}
\begin{subfigure}[tb]{\columnwidth}
\centerline{\includegraphics[width=0.93\columnwidth, trim=10mm 0mm 24mm 13mm, clip]{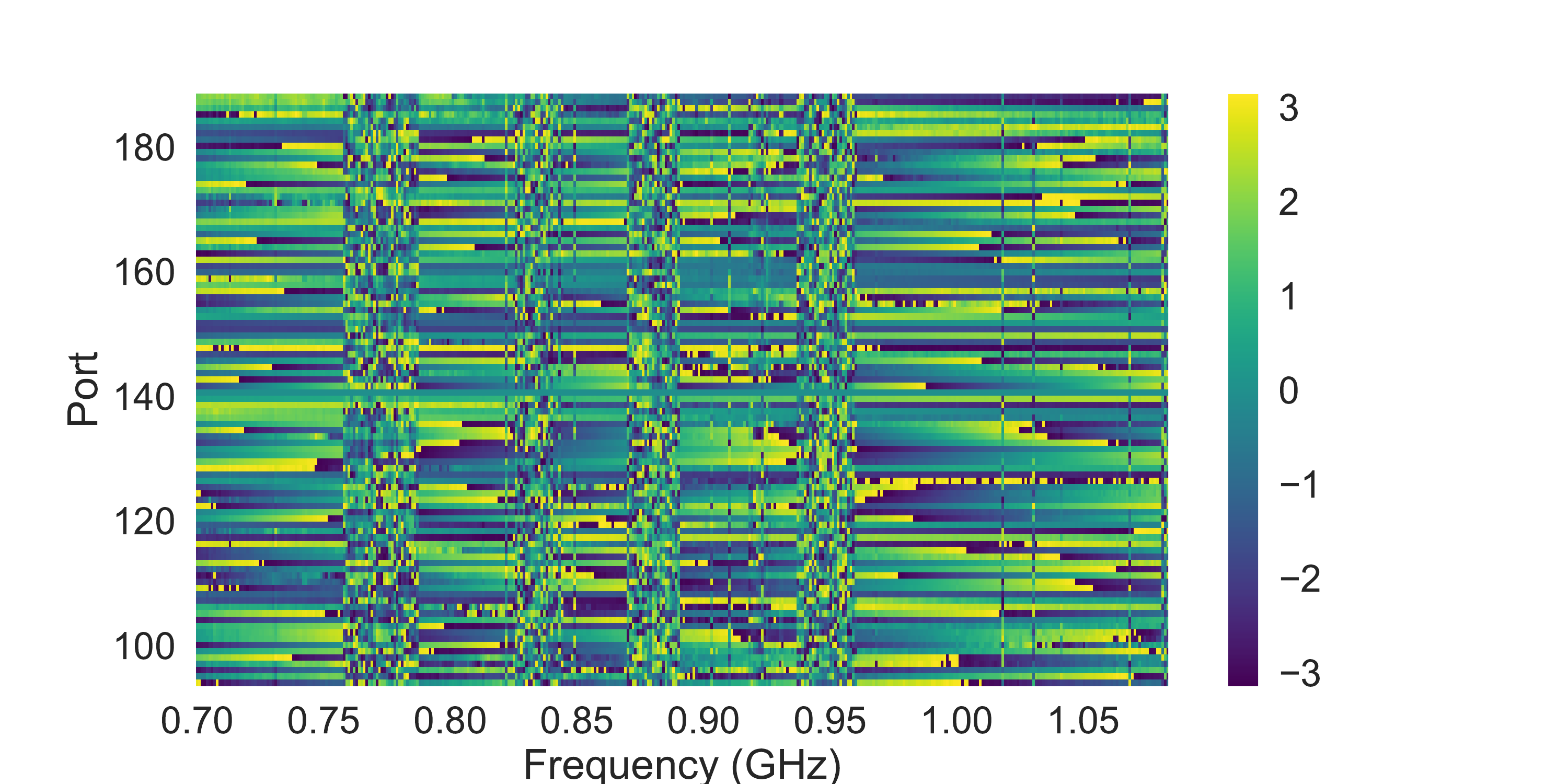}} 
\caption{\small Raw.\label{fig:phase} }
\end{subfigure}

\begin{subfigure}[tb]{\columnwidth}
\centerline{\includegraphics[width=0.93\columnwidth, trim=10mm 0mm 24mm 13mm, clip]{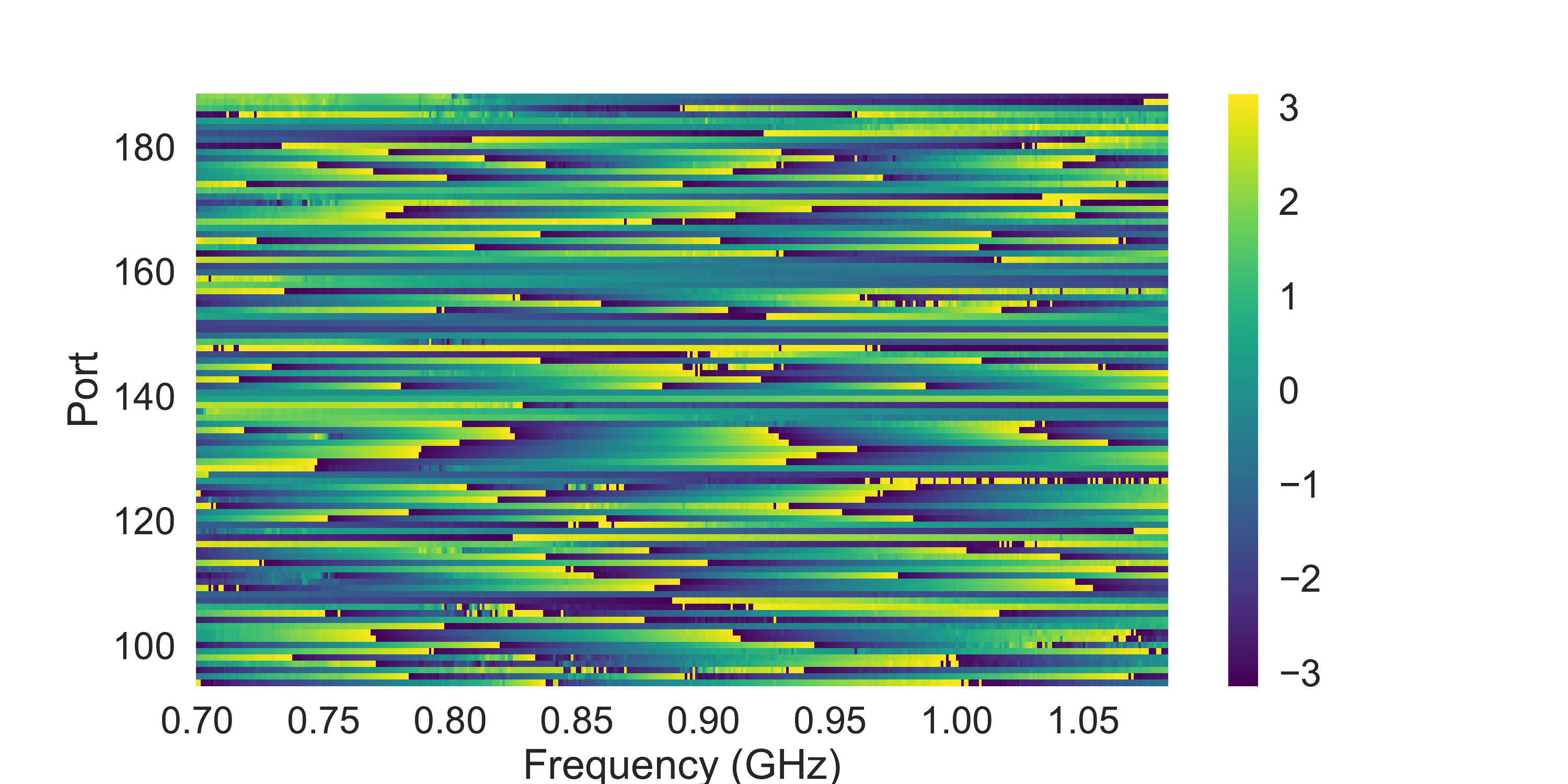}}
\caption{\small Interpolated.\label{fig:phase-interp} }
 \end{subfigure}
 \caption{MaxSNR beam weight phases (radians).}
\end{figure}

Correct initial beam weights, that maximise the desired signal, are required for algorithms that suppress RFI and retain the desired signal.  Figures \ref{fig:amp} and \ref{fig:phase} show that the initial maxSNR beam weights are corrupted at some frequencies by RFI. Here all weights were normalised by
%dividing through by 
the weights of the port with the highest typical weight amplitude\footnote{PAF port 141 in the numbering of ACES Memo 001 at \url{http://www.atnf.csiro.au/projects/askap/ACES-memos}.}.  %This makes the weight of that reference port unity amplitude and zero phase at all frequencies.  

Figures \ref{fig:amp-interp} and \ref{fig:phase-interp} show that we can recover the weights at interference-affected channels.  Amplitude is interpolated independently for each port by iteratively fitting a polynomial to its weight amplitude as a function of frequency and removing outliers that are likely to be interference-affected channels at each iteration.  Before fitting a polynomial to the weight phase of a given port, we first removed a bulk linear phase (delay) estimated by taking the Fourier transform of the normalised weights with respect to frequency.  We then unwrapped the phase and iteratively fit a polynomial to interference-free channels as described above.  In Figure \ref{fig:phase-interp}, weights corrupted by RFI are replaced with an evaluation of the fitted polynomials plus the bulk linear phase.

\section{CONCLUSION}
We have suppressed RFI by up to \SI{20}{\decibel}  in real-time observations with a PAF on the Parkes telescope.  Suppression could be improved by reducing the integration time and increasing the download rate of ACMs to account for dynamic RFI environments, and by correlating PAF ports against reference antennas directed at RFI sources to enhance subspace estimation.  Within current limitations, best performance was achieved on a narrowband and stationary clock signal originating from the PAF back-end.  This was cleanly mitigated to the noise floor with no reduction in system sensitivity.  Finally, we demonstrated successful interpolation of beam weights at interference-affected channels. This is immediately useful for general PAF beamforming in addition to being necessary for mitigation algorithms that robustly preserve the desired signal.

\section*{Acknowledgments}
{\small The Parkes radio telescope is part of the ATNF, which is funded by the Commonwealth of Australia for operation as a National Facility managed by the CSIRO. MPIfR financed the PAF used in this paper and its modification for a less radio-quiet site.  Dr. K. Bannister and C. Haskins supported software implementation.  A. Brown improved the duty cycle of the FFB integrator.  Dr. B. Inderm\"uhle supported signal identification.  Dr. J. Tuthill advised on the digital receiver read-out clock. Prof. L. Staveley-Smith motivated this work. }
\bibliographystyle{IEEEtran}
\bibliography{IEEEabrv,apj-jour,intmit,aces-memos}

\end{document}